\documentclass[a4paper]{article}
\usepackage{spconf,amsmath,epsfig,graphics,url, subcaption}
\usepackage{hyperref}

\usepackage{fancyhdr}
\title{Simulation of Plenoptic Cameras}

\name{Tim Michels, Arne Petersen, Luca Palmieri, Reinhard Koch}
\address{CAU Kiel, Department of Computer Science}

\begin{document}
\ninept
\urlstyle{same}
\maketitle
\fancyfoot[C]{\footnotesize{$\copyright$ 2018 IEEE. Personal use of this material is permitted.
		Permission from IEEE must be obtained for all other uses, in any current or future
		media, including reprinting/republishing this material for advertising or promotional
		purposes, creating new collective works, for resale or redistribution to servers or
		lists, or reuse of any copyrighted component of this work in other works.
		DOI: \href{https://doi.org/10.1109/3DTV.2018.8478432}{10.1109/3DTV.2018.8478432}}}

\begin{abstract}
	Plenoptic cameras enable the capturing of spatial as well as angular color information which can be used for various applications among which are image refocusing and depth calculations. However, these cameras are expensive and research in this area currently lacks data for ground truth comparisons. 
	In this work we describe a flexible, easy-to-use Blender model for the different plenoptic camera types which is on the one hand able to provide the ground truth data for research and on the other hand allows an inexpensive assessment of the cameras usefulness for the desired applications. Furthermore we show that the rendering results exhibit the same image degradation effects as real cameras and make our simulation publicly available.
\end{abstract}


\section{Introduction}
\begin{figure}[!t]
	\centering
	\includegraphics[width=.41\textwidth]{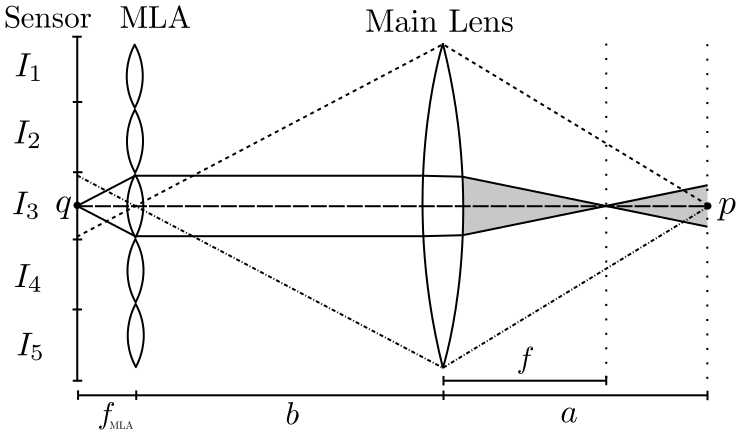}
	\caption{Plenoptic 1.0 camera as introduced by Adelson and Wang \cite{adelsonwang1992plenopticcam} and implemented by Ng \cite{ng2005lightfieldcamera}: The MLA is focused at infinity. Given $\frac{1}{f}=\frac{1}{a}+\frac{1}{b}$ the scene point $p$ is seen by multiple pixels of the microlens image $I_3$ from slightly different angles. These pixels, however, also see a certain area around $p$ as exemplarily shown for the pixel $q$, which results in a high angular but low spatial resolution \cite{perwass2012single}.}
	\label{fig:plenoptic1}
\end{figure}

\begin{figure}[!t]
	\centering
	\includegraphics[width=.41\textwidth]{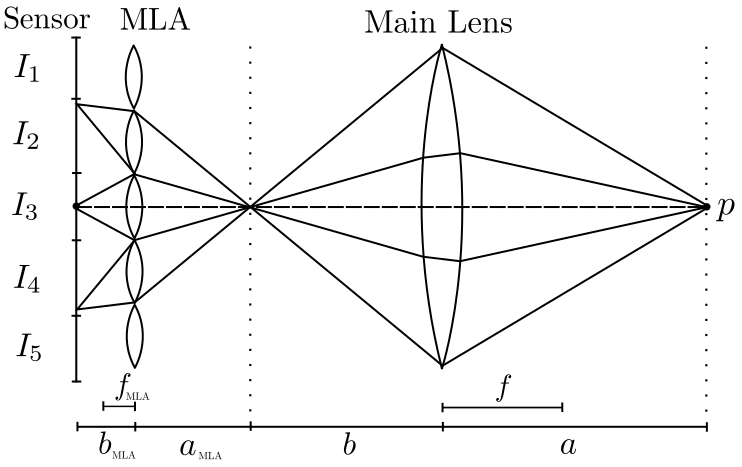}
	\caption{Plenoptic 2.0 camera as introduced by Lumsdaine and Georgiev \cite{lumsdaine2009focusedplenoptic}: The MLA is focused on the virtual image of the main lens. Given $\frac{1}{f}=\frac{1}{a}+\frac{1}{b}$ for the main lens as well as the microlenses, the scene point $p$ is seen by multiple microlenses, but only one pixel per microlens image. This results in higher spatial but lower angular resolution compared to plenoptic 1.0 cameras \cite{perwass2012single}.}
	\label{fig:plenoptic2}
\end{figure}

Tracing back to the ideas of Lippmann \cite{lippmann1908integralphoto} and Ives \cite{ives1928camera} the concept of capturing light fields with a single camera has regained interest during the past decade due to its commercially available realizations in the form of plenoptic cameras by Lytro \cite{lytro} and Raytrix \cite{raytrix}. Depending on the model, these cameras can be rather expensive, thus an accurate simulation of plenoptic cameras could enable a cheap and uncomplicated assessment of the usefulness for different applications. Furthermore, major parts of the research related to plenoptic cameras are focused on the their calibration and the reconstruction of depth images as well as color images with a modified depth of field (DoF) or altered viewpoint. These would greatly benefit from realistic, simulated ground truth data. However, the simulation of realistic plenoptic camera data is non-trivial due to the setup of these cameras. The basic concept of the two types of plenoptic cameras shown in \autoref{fig:plenoptic1} and \autoref{fig:plenoptic2} is the use of a microlens array (MLA) between a conventional camera's main lens and its image sensor in order to capture not only position dependent but also view angle dependent information. Accordingly every light ray that reaches the sensor of a plenoptic camera has passed through the main lens system and one microlens and thus is affected by the properties of both.\\
While several related works on light field imaging focus on camera array data, the works that actually include or explicitly describe the simulation of plenoptic cameras usually ignore some part of the multi-lens setup leading to unrealistically perfect data. Fleischmann et al. \cite{fleischmann2014plenoptic} synthesize plenoptic camera data without using a main lens, thus reducing the setup to a simple multi-camera array. Zhang et al. \cite{zhang2015forwardsimulation} and Liang et al. \cite{liang2015simuWOmainlens} use a simplified thin main lens, which does not lead to the degradation effects visible in real images, and Liu et al. \cite{liu2015numericalsimulation} require the captured scene objects to be at an unrealistically large distance from the camera. In addition most of these previous works use forward ray tracing and restrict themselves to scene objects with simple geometries and Lambertian surfaces.\\
Our contribution is a physically-based simulation of plenoptic 1.0 and 2.0 cameras in Blender \cite{blender} which includes a realistic model of the main lens as well as a configurable MLA and is made publicly available\footnote{\href{https://github.com/Arne-Petersen/Plenoptic-Simulation}{https://github.com/Arne-Petersen/Plenoptic-Simulation}}. Furthermore we analyze our synthesized images and show that these exhibit similar geometric and photometric degradation effects as images from Raytrix or Lytro cameras.


\section{Lens Effects in Plenoptic Cameras}
\begin{figure}[!t]
	\centering
	\includegraphics[width=.49\textwidth]{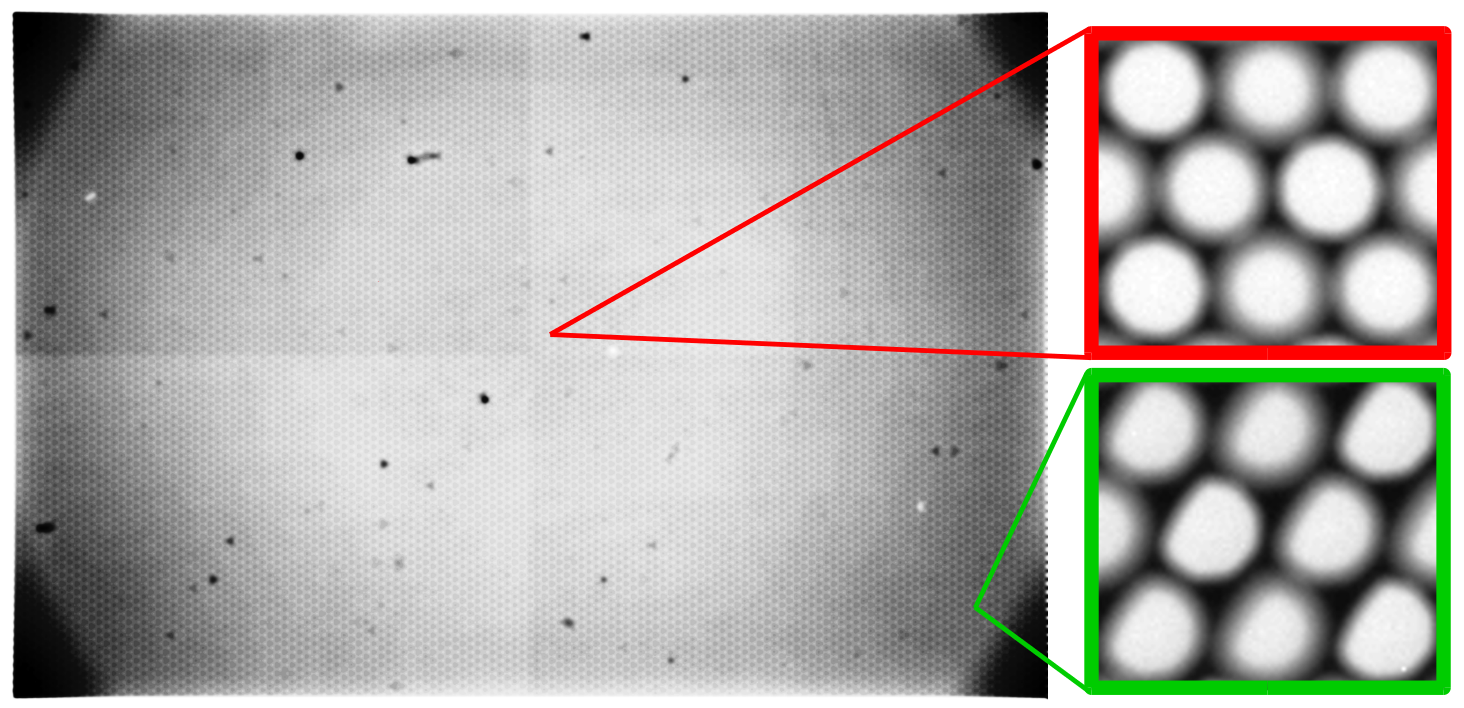}
	\caption{Vignetting effects: The left picture shows a heavily contrast enhanced image from a Raytrix R29 capturing an evenly illuminated white plane. Bright and dark spots are a result of microlens imperfections or dust particles on the MLA in combination with the contrast enhancement. The right pictures are unaltered sections of the original image showing the effect of the main lens aperture on the microlens images.}
	\label{fig:vignetting}
\end{figure}

In this section we will give a description of the photometric and geometric degradation effects in plenoptic cameras. Due to the combination of the main and microlenses the effects that are observable in standard cameras are also present in plenoptic cameras, but have a different impact on the image quality. In the following we list major degradation effects and the problems arising in synthesizing them after rendering a defect-free multi-camera image without explicitly modeling the lenses of a plenoptic camera.\\
\textit{Radial distortion} affects the main lens as well as the microlenses. However, the radial distortion of the microlenses is neither significant, because of the low resolution of the microlens images, nor efficient to handle due to the high number of microlenses. On the contrary, the radial distortion of the main lens plays a significant role in the plenoptic imaging process \cite{johannsen2013calibration}.
Synthesizing this effect via a simple inverse application of the radial distortion polynomial would result in a shift of the microlens images on the sensor leading to incorrect correspondences between a microlens area on the sensor and its image position.\\
\textit{Vignetting} in the final image results from the addition of vignetting effects from the main lens and microlenses as shown in \autoref{fig:vignetting}.
The main lens vignetting, influenced by its aperture, does not only affect the amount of light reaching the microlenses, it also defines the shape of a microlens image. Furthermore, with increasing distance from the main lens optical axis, the microlens images are cut off to one side (compare \autoref{fig:vignetting}) due to a limited exit pupil. These effects of the main lens vignetting are further amplified by the microlens aperture which causes additional vignetting. 
Despite the correction of this vignetting being simple via classical white image division \cite{yu2004practical}, the complexity of the combined effect poses a problem for the feasibility of its synthesis. In order to generate the correct vignetting for a certain camera configuration a complex model is needed taking into account the main lens aperture configuration as well as the microlens position with respect to the main lens optical axis.\\
\textit{Depth distortion} describes the influence of the microlens distance to the main lens optical axis on depth reconstruction algorithms. With increasing distance the depth error increases as a result of the so-called Petzval field curvature \cite{johannsen2013calibration}. In order to synthesize this effect in a perfect image for a certain objective a model similar to the radial distortion has to applied. This however requires precise knowledge on the Petzval field curvature of the given lenses.\\
\textit{Coma} and \textit{astigmatism} are further effects which, like the Petzval field curvature, can influence the focal properties of the lens and thereby the depth reconstruction. Using modern objectives their effects on the final image are usually negligible and accordingly it is only necessary to synthesize these effects if the main lens is known to exhibit a significant amount of these distortions. Then, however, the synthesis poses the same problem as the Petzval field curvature, namely the necessity of measuring these effects in order to build a precise distortion model.\\
Apart from these degradation effects the main lens also alters the scene geometry seen by the MLA. The virtual scene, given by projecting the real scene through the main lens, is a non-linearly scaled version of the real scene. While the geometry of the virtual scene could approximately be calculated by applying the thin or thick lens equation to the real scene, the textures and especially the lighting are hard to synthesize since straight rays of the real scene, as usually used in ray tracing applications, are projected to curves in the virtual image. Accordingly the use of a main lens model significantly reduces the complexity of the virtual scene formation.\\
Further types of degradation are given by the imperfections of real cameras such as lens material defects, dead pixels, inaccurately mounted objectives or simply dust particles on lenses or sensors. Depending on the extent of the defect or inaccuracy the effect ranges from the degradation of single microlenses (see \autoref{fig:vignetting}) to complex \textit{caustics} or a \textit{tangential distortion} of the whole image due to a main lens tilt. Like most of the previously mentioned effects, these can also be synthesized with varying expenditure. Nevertheless, a combination of all or a subset of these effects would require several and in some combinations even more complex distortion models since some of the effects depend on each other. Therefore the easiest solution to overcome the necessity of unfeasible distortion models is the direct use of accurately modeled main and microlenses as described in the following section.

\section{Blender Modeling}
\begin{figure}[!t]
	\centering
	\includegraphics[width=.4\textwidth]{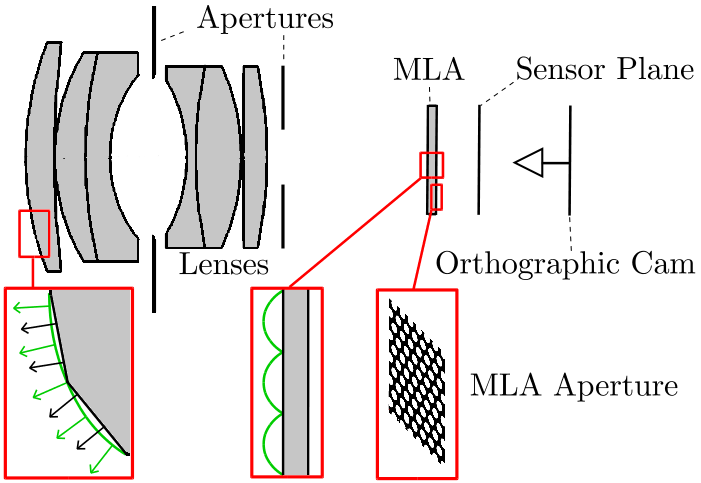}
	\caption{Overview of our plenoptic camera model in Blender including a comparison of real geometry (green) and the approximation in Blender (black).}
	\label{fig:overviewblender}
\end{figure}
The use of Blender for modeling a plenoptic camera has several advantages. As a free and widely used software it enables every interested person to easily create realistic images by using the implemented Cycles render engine. This renderer is physically-based and supports different path tracing variants which allow object models with refractive materials to exhibit nearly the same effects as their real pendants. Therefore Blender is not only suitable to simulate multiple lenses but, in contrast to previous works' simulations, also allows the use of complex scenes including non-Lambertian and refractive materials as well as complex lighting setups. However, Blender also has some limitations regarding the overall number of vertices and minimal correctly simulatable size of objects and distances between them, posing different challenges for the modeling of a plenoptic camera.\\
While the objective's lenses can theoretically be modeled as sphere intersections, the limitation in the number of vertices leads to lens models with only approximately round surfaces. When such a model and its surface normals are used for the rendering, the final image shows triangle shaped artifacts resulting from the discretization of the surface and its normals. However, since the correct lens geometry can be seen as a combination of the approximating 3D model and additional thin lenses added to its surface (see \autoref{fig:overviewblender}), it suffices to only calculate the correct surface normals in order to simulate the expected lens behavior. A geometry correction is not necessary because thin lenses can be approximated by simple refracting planes. In other words, slightly incorrect surface normals distort the projection of rays through lenses significantly more than marginal surface displacements. Therefore we simply use the lens models material shader to calculate the correct surface normals and overcome this discretization issue.\\
In addition to the lenses and the objective's aperture we add another aperture at the objective's exit (see \autoref{fig:overviewblender}) to simulate the limited exit pupil as discussed in the context of vignetting.\\
Since the microlens effects are mostly negligible compared to the main lens effects it seems natural to render each microlens image separately by using the internal Blender camera and shifting it to the next microlens position afterwards. However, the possible Blender camera settings are bounded below in respect of focal distance, sensor size and DOF values. This leads to a restricted MLA configuration and e.g. prevents realistic results when the MLA is placed between the main lens and the virtual image which is a common setup for plenoptic 2.0 cameras. Therefore it is necessary to explicitly model the MLA as well as the image sensor. Because of the limitation regarding the total number of vertices it is impossible for the microlenses to be realistically modeled since MLAs usually consist of up to $2\cdot 10^5$ microlenses each of which would need a smooth, round surface. Since we are mainly interested in creating microlens models with the correct focal length, we can use the lensmaker's equation, which states the exact relation between focal length, radius of the front surface of the microlens and the IOR, in order to find a simpler MLA model. For a lens with IOR $n$, front surface curvature radius $R_1$, a small thickness $d\approx 0$ and a flat back surface, i.e. $R_2=\infty$, as shown in \autoref{fig:overviewblender}, we get 
\begin{equation*}
\frac{1}{f}=(n-1)\left(\frac{1}{R_1}-\frac{1}{R_2}+\frac{(n-1)d}{nR_1R_2}\right)=\frac{n-1}{R_1}.
\end{equation*}
Accordingly, the same focal length $f$ can be achieved with every thin lens satisfying $R_1/(n-1)=f$, especially lenses with nearly flat front surfaces and high IOR. As previously mentioned, the normals and index of refraction (IOR) are more important for the correct refraction of a ray than the exact surface geometry. Therefore a microlens with a nearly flat front surface can be approximated by a flat lens with recalculated normals. Hence we use a simple two plane model with high IOR for the MLA and calculate the correct normals for the nearly flat lenses in the MLA model's material shader. Furthermore, we mask the back surface of the MLA to simulate the microlens apertures (compare \autoref{fig:overviewblender}).\\
Finally the image sensor of the plenoptic camera is simulated as a combination of a simple plane equipped with a refractive shader and an orthographic Blender camera viewing the plane. While a real camera sensor pixel has a certain FOV and collects light from this range of directions, a perfectly refractive plane refracts a camera viewing ray into only one exact direction and therefore the corresponding orthographic camera pixel only sees a fraction of the light reaching the sensor plane. Thus, roughness is added to the refraction shader in order to allow the camera viewing rays for one pixel to be refracted randomly within a range of slightly different directions and therefore accumulating a realistic amount of light.

\section{Evaluation}
\begin{figure}[!t]
	\centering
	\includegraphics[width=.48\textwidth]{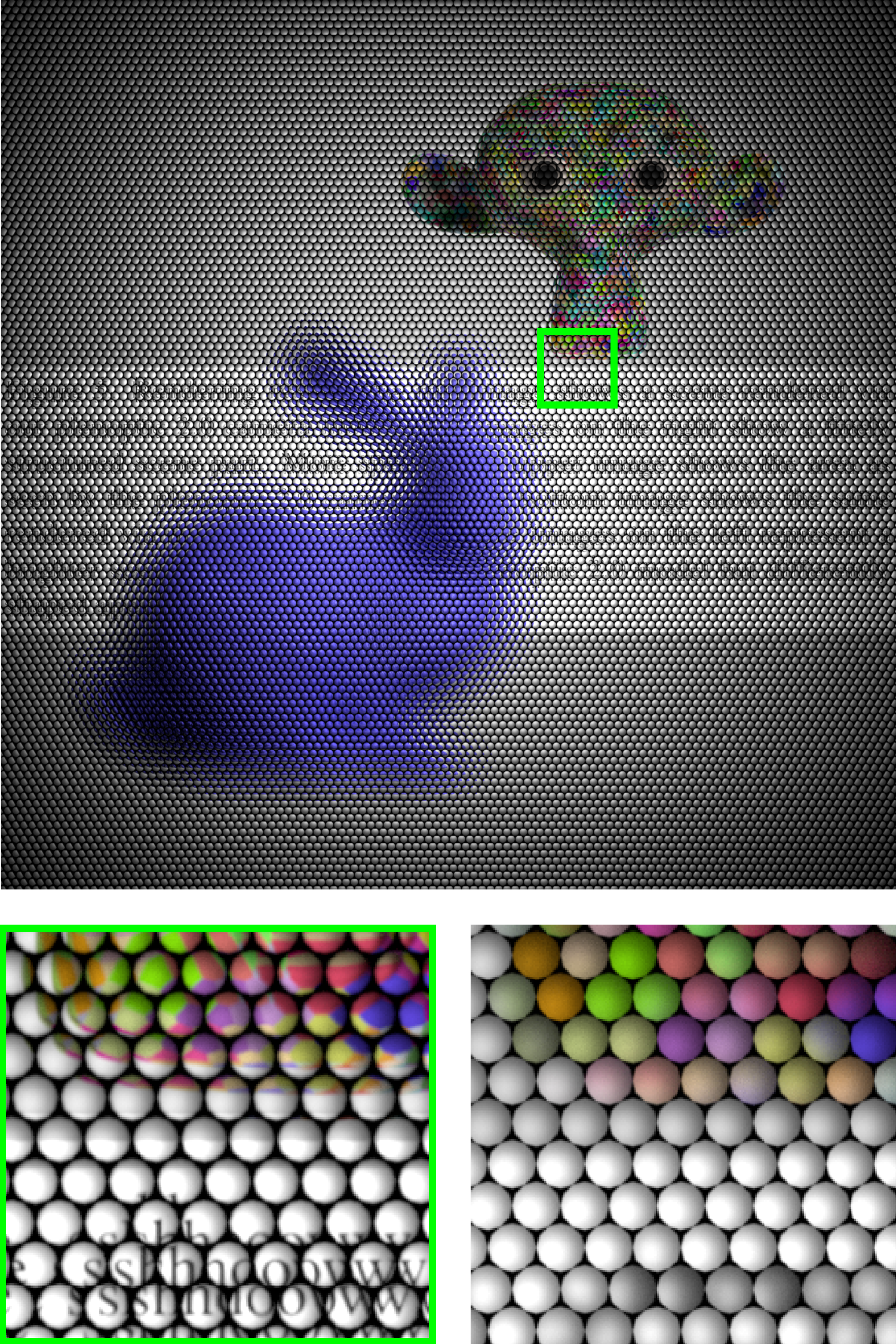}
	\caption{Top: A scene rendered via our plenoptic 2.0 camera model. Bottom: Comparison of a finely structured scene part, marked green in the top image, rendered via the plenoptic 2.0 (left) and 1.0 (right) model.}
	\label{fig:result_comparison}
\end{figure}
\begin{figure}[!t]
	\centering
	\includegraphics[width=.22\textwidth]{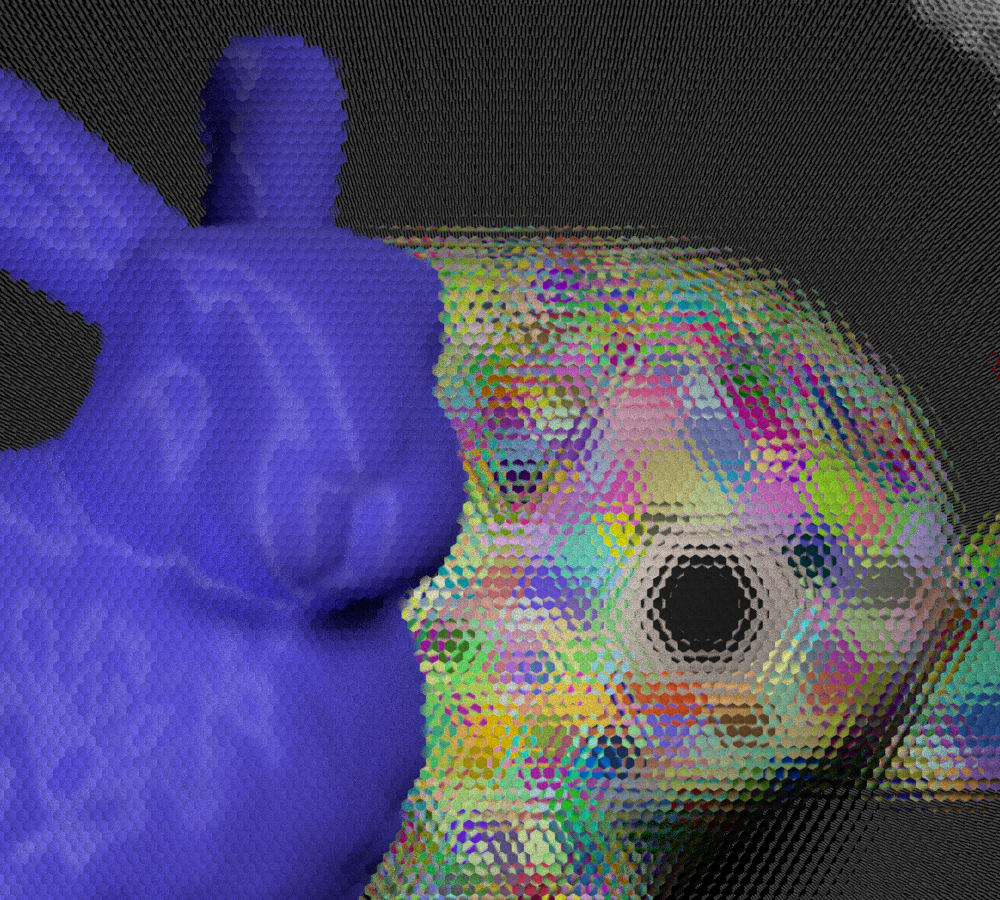}
	\quad
	\includegraphics[width=.22\textwidth]{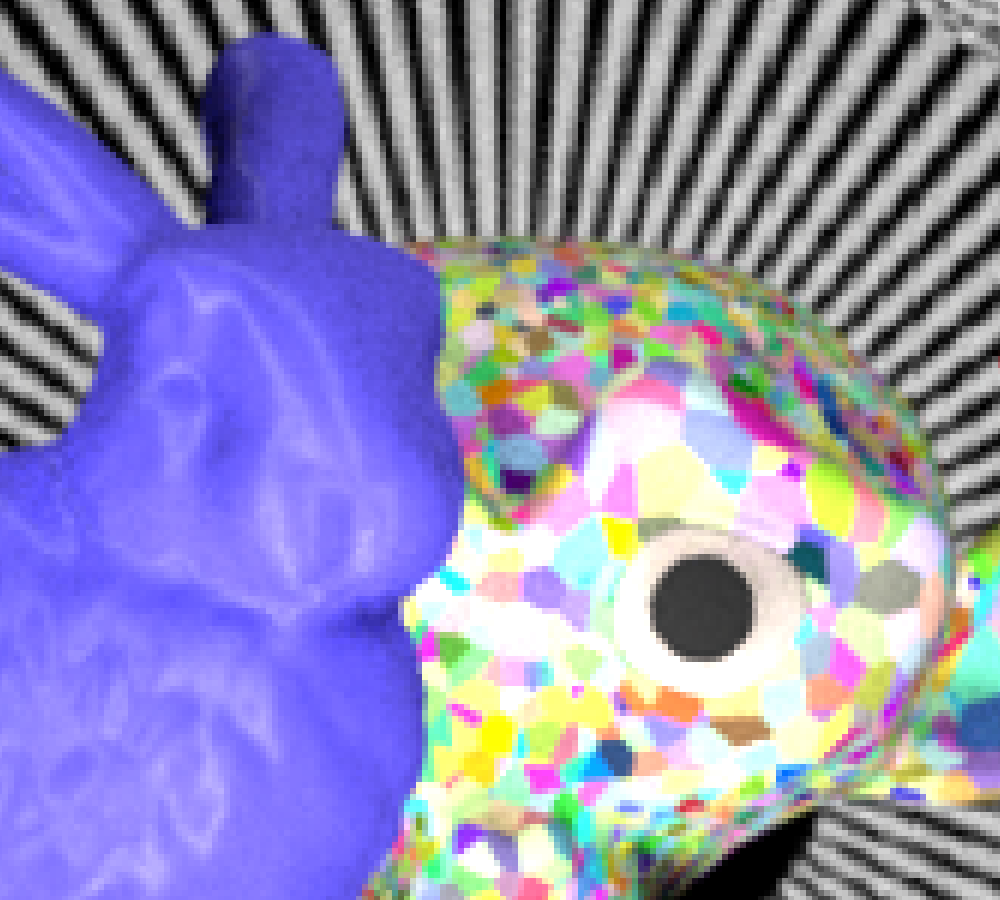}
	\caption{Cutout of a plenoptic 1.0 rendering and the corresponding section of one sub-aperture image.}
	\label{fig:result_subap}
\end{figure}
\begin{figure}[!t]
	\centering
	\includegraphics[width=.14\textwidth]{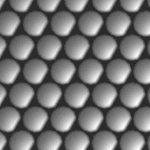}
	\quad
	\includegraphics[width=.14\textwidth]{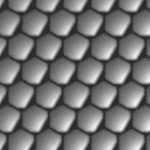}
	\quad
	\includegraphics[width=.14\textwidth]{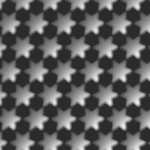}
	\caption{Vignetting effects in microlens images resulting from differently shaped (12-blade, 6-blade and star-shaped) main lens apertures and a limited exit pupil.}
	\label{fig:result_vignetting}
\end{figure}

For our tests we constructed a 100mm objective according to the double gaussian lens model described in \cite{kolb1995lensmodelforsimulation}. Furthermore we used microlenses with a focal length of 2mm and a diameter of 0.217mm. For the plenoptic 2.0 camera setup, the MLA distance to the objectives center was set to 123.3mm and the distance between MLA and sensor plane to 1.7mm. Consequently the microlens focal points are located slightly behind the sensor plane or, from a different point of view, the MLA is not focused at infinity but to a distance of 11.33mm according to the thin lens equation. For the plenoptic 1.0 setup, on the contrary, the sensor plane is placed exactly 2mm behind the MLA, thus setting its focal distance to infinity.\\
Here we would like to remark, that our results slightly differ from the images captured with real Raytrix cameras as shown in \autoref{fig:vignetting} since we only use one microlens type and the modeled objective is not equal to the real 100mm objective used for capturing that image. Moreover the extent to which the previously discussed effects are observable heavily depends on the objective. In our case the modeled objective exhibits radial as well as depth distortion to an extent that is only measurable but has nearly no visible effect on the renderings. These distortions, however, are inherent properties of lenses simulated via ray tracing thus there is no need for further validation regarding this aspect. Nevertheless, the rendering results show some other effects that are not only measurable but also clearly visible. As shown in \autoref{fig:result_comparison}, the renderings from the different plenoptic camera types exhibit the expected differences with respect to the trade-off between angular and spatial resolution for objects at a distance $a$ that approximately satisfies the thin lens equation as shown in \autoref{fig:plenoptic1} and \autoref{fig:plenoptic2}. While the plenoptic 2.0 images preserve fine details of the scene, the corresponding plenoptic 1.0 microlens images show a significant higher amount of blur.
These correct imaging properties regarding the geometry can be further verified by the sub-aperture images (as exemplarily shown in \autoref{fig:result_subap}) which exhibit the expected occlusion and reflection behavior. Here, due to the limited space, we refer the reader to our aforementioned repository, where additional renderings are available that show these effects.\\
Finally, the expected vignetting is also clearly observable. The rendered images show the main lens vignetting in regards of decreasing brightness towards the sensor edges (see \autoref{fig:result_comparison}) as well as regarding the shape and cut-off effect of the single microlens images (see \autoref{fig:result_vignetting}).

\section{Conclusion}
While our model simulates plenoptic camera data quite realistically, there is still some room for improvement. Images of real (plenoptic) cameras exhibit chromatic aberrations and artifacts related to debayering as well as imperfections in the microlenses
. Furthermore a variety of different main lens models could be implemented in order to simulate more complex and more recent objectives than the simple double gaussian lens model we used. Nevertheless, with our simulation we present a useful basis for future research that already covers the majority of the otherwise hard to synthesize effects.

\section*{Acknowledgment}
This work was supported by the German Research Foundation, DFG, No. K02044/8-1 and the EU Horizon 2020 program under the Marie Sklodowska-Curie grant agreement No 676401.

\bibliographystyle{IEEEtran}
\bibliography{bibliography/references}

\end{document}